\documentclass[11pt]{article}

\usepackage[utf8]{inputenc}
\usepackage[english]{babel}

\usepackage{hyperref}
\usepackage{xcolor}
\usepackage{amsmath,amssymb}
\usepackage{enumerate}
\usepackage{amsthm}
\usepackage[all,2cell]{xy}
\usepackage{mathrsfs}
\usepackage{mathtools}
\usepackage{tikz}

\usepackage{geometry}
\newtheorem{thm}{Theorem}[section]
\newtheorem{dfn}[thm]{Definition}
\newtheorem{prop}[thm]{Proposition}

\def\beq{\begin{equation}}
\def\eeq{\end{equation}}
\def\bea{\begin{eqnarray}}
\def\eea{\end{eqnarray}}
\def\beann{\begin{eqnarray*}}
\def\eeann{\end{eqnarray*}}
\def\ben{\begin{enumerate}}
\def\een{\end{enumerate}}
\def\bit{\begin{itemize}}
\def\eit{\end{itemize}}

\newcommand\restr[2]{{
  \left.\kern-\nulldelimiterspace 
  #1 
  \right|_{#2} 
}}

\title{\sc
A contact geometry approach to symmetries in systems with dissipation}

\author{\sffamily 
Jordi Gaset 
\thanks{emails: 
jordi.gaset@uab.cat}
\\[1ex]
\normalsize\itshape\sffamily 
Department of Physics,
Universitat Aut\`onoma de Barcelona,
}

\date{\sffamily  October, 2019}

\begin{document}

\maketitle

\begin{abstract}
Systems with dissipation can be described using contact geometry. We introduce the concepts of symmetries and dissipation laws for contact Hamiltonian systems and study the relation between them.  This is an ongoing collaboration with Xavier Gr\`acia, Miguel C. Mu\~noz-Lecanda,  Xavier Rivas and Narciso Rom\'an-Roy.
\end{abstract}

\section{Introduction}

In  many  mechanical  systems  without  dissipation,  we  are  interested  in  quantities (like energy or the different momenta)  which  are conserved along a solution. They are an effective tool to understand and integrate the system. From a physical point of view, if a system has dissipation, these quantities are not conserved. Since damped systems rarely have a standard Lagrangian or Hamiltonian formulation, this problem can not be studied with the usual tools.

There is a growing interest in describing the geometrical framework of dissipated or damped systems, specifically using contact geometry \cite{bravetti,scontact,LeonLainz, LeonLainz2}. All of them are described by ordinary differential equations to which some terms that account for the dissipation or damping have been added. In order to provide a variational formulation for these systems, contact geometry introduces a new variable or parameter, together with a new set of equations. It turns out that this variable is closely related to the action itself, and some authors consider these theories as described by an action-dependent Lagrangian. Contact Hamiltonian systems provide us the geometric framework we will use to analyse symmetries and (non)-conserved quantities.

First we will present the geometric structures of the contact formalism, and contact Hamiltonian systems. Then we will define several classes of symmetries for this kind of systems, which have different properties. 

The analogous concept of conserved quantities are called dissipated quantities. In the contact formalism, the evolution of these quantities is determined by a dissipation law. We will also show how to construct conserved quantities from dissipated quantities.

It is possible to relate symmetries with dissipated quantities, in a result inspired in Noether's Theorem. We will briefly discuss how this relation works. Finally, we will apply this tools to the motion in a gravitational field with friction.

This framework can be extended to describe field theories with dissipation introducing the concept of $k$-contact structures \cite{kcontact}.



\section{Contact  Manifolds and Contact Hamiltonian Systems}
			
\begin{dfn}\label{dfn-contact-manifold}
    Let $M$ be a $(2n+1)$-dimensional  manifold. 
    A \textbf{contact form} in $M$ is a differential $1$-form
    $\eta\in\Omega^1(M)$ such that $\eta\wedge(\mathrm{d}\eta)^{\wedge n}$
    is a volume form in $M$.
    Then, $(M,\eta)$ is said to be an  \textbf{ (exact) contact manifold}.
\end{dfn}

As a consequence of the condition that
$\eta\wedge(\mathrm{d}\eta)^{\wedge n}$ is a volume form we have a decomposition of 
$\mathrm{T} M$, induced by $\eta$, in the form 
$\mathrm{T} M= \ker\mathrm{d}\eta\oplus\ker\eta$. Therefore, there exists a unique vector field $\mathcal{R}\in\mathfrak{X}(M)$, 
which is called \textbf{Reeb vector field}, 
such that
\begin{equation}\label{eq-Reeb}
    \begin{cases}
        \mathrm{i}(\mathcal{R})\mathrm{d}\eta = 0,\\
        \mathrm{i}(\mathcal{R})\eta = 1.
    \end{cases}
\end{equation}

This vector field generates the distribution $\ker \mathrm{d} \eta$, 
which is called the {\sl Reeb distribution}. In a contact manifold one can prove the existence of Darboux-type coordinates:

\begin{thm}[Darboux theorem for contact manifolds]
Let $(M,\eta)$ be a contact manifold. 
Then around each point $p\in M$ there exist a chart 
$(\mathcal{U}; q^i, p_i, s)$ with $1\leq i\leq n$ such that
\begin{equation*}
	{\eta}|_{\mathcal{U}} = \mathrm{d} s - p_i\,\mathrm{d} q^i \  .
\end{equation*}
These are the so-called \textbf{Darboux} or \textbf{canonical coordinates} of the contact manifold $(M,\eta)$.
\end{thm}

In Darboux coordinates, the Reeb vector field is
$\displaystyle{\mathcal{R}}|_{\mathcal U} = \frac{\partial}{\partial s}$.

\begin{thm}
\label{teo-hameqs}
    If $(M,\eta)$ is a contact manifold, for every $\mathcal{H}\in\ C^\infty(M)$,
     there exists a unique vector field $X_\mathcal{H}\in\mathfrak{X}(M)$ such that
    \begin{equation}\label{hamilton-contact-eqs}
        \begin{cases}
            \mathrm{i}(X_\mathcal{H})\mathrm{d}\eta=\mathrm{d}\mathcal{H}-(\mathrm{L}_{\mathcal{R}}\mathcal{H})\eta\\
            \mathrm{i}(X_\mathcal{H})\eta=-\mathcal{H} \ .
        \end{cases}
    \end{equation}
\end{thm}

The vector field $X_\mathcal{H}$ is the
\textbf{contact Hamiltonian vector field} associated to $\mathcal{H}$ and the equations 
\eqref{hamilton-contact-eqs} are the \textbf{contact Hamiltonian equations} 
for this vector field. The triple $(M,\eta,\mathcal{H})$ is a \textbf{contact Hamiltonian system}.

As a consequence of the definition of $X_\mathcal{H}$ we have
the following relation,
which expresses the dissipation of the Hamiltonian:
$$
\mathrm{L}_{X_\mathcal{H}}\mathcal{H} =
-(\mathrm{L}_{\mathcal{R}}\mathcal{H})\mathcal{H} \:.
\label{eq:disipenerg}
$$
Indeed: $
\mathrm{L}_{X_\mathcal{H}}\mathcal{H}=-\mathrm{L}_{X_\mathcal{H}}\mathrm{i}(X_\mathcal{H})\eta=
-\mathrm{i}(X_\mathcal{H})\mathrm{L}_{X_\mathcal{H}}\eta=
\mathrm{i}(X_\mathcal{H})((\mathrm{L}_\mathcal{R}\mathcal{H})\eta)=-(\mathrm{L}_{\mathcal{R}}\mathcal{H})\mathcal{H} \ .
$

Taking Darboux coordinates $(q^i,p_i,s)$, 
the contact Hamiltonian vector field is
$$ 
X_\mathcal{H} = \frac{\partial\mathcal{H}}{\partial p_i}\frac{\partial}{\partial q^i} - 
\left(\frac{\partial\mathcal{H}}{\partial q^i} + 
p_i\frac{\partial\mathcal{H}}{\partial s}\right)\frac{\partial}{\partial p_i} + 
\left(p_i\frac{\partial\mathcal{H}}{\partial p_i} - \mathcal{H}\right)\frac{\partial}{\partial s}\ ; 
$$
and its integral curves $\gamma(t) = (q^i(t), p_i(t), s(t))$
are solutions of
\begin{equation}\label{diss-ham-eqs}
\begin{cases}
    \dot q^i = \frac{\partial\mathcal{H}}{\partial p_i}\ ,\\
    \dot p_i = -\left(\frac{\partial\mathcal{H}}{\partial q^i} + p_i\frac{\partial\mathcal{H}}{\partial s}\right)\ ,\\
    \dot s = p_i\frac{\partial\mathcal{H}}{\partial p_i} - \mathcal{H}\ .
\end{cases}
\end{equation}

\section{Symmetries and Dissipation Laws for Contact Hamiltonian Systems}

One can consider different concepts of symmetry in a dynamical system, which depend on which structure they preserve. We will define dynamical symmetries, which preserve the space of solutions, and contact symmetries, which preserve the geometric structure.

Let $(M,\eta,\mathcal{H})$ be a contact Hamiltonian system with Reeb vector field $\mathcal{R}$,
and $X_\mathcal{H}$ the contact Hamiltonian vector field for this system;
that is, the solution to the Hamilton equations \eqref{hamilton-contact-eqs}.

\begin{dfn}\label{dfn:sym} Consider a diffeomorphism $\Phi\colon M\longrightarrow M$ and a vector field $Y\in \mathfrak{X}(M)$:
\begin{itemize}
    \item $\Phi$ is a \textbf{dynamical symmetry} if $\Phi_*X_\mathcal{H}=X_\mathcal{H}$
    (it maps solutions into solutions). $Y$ is an \textbf{infinitesimal dynamical symmetry} if its local flows are dynamical symmetries; that is,
    $\mathrm{L}_YX_\mathcal{H}=[Y,X_\mathcal{H}]=0$.
    
    \item $\Phi$ is a \textbf{contact symmetry} if
    $$
        \Phi^*\eta=\eta
        \quad ,\quad \Phi_*\mathcal{H}=\mathcal{H}\ .
    $$    
    $Y$ is an \textbf{infinitesimal contact symmetry} if its local flows are contact symmetries; that is,
    $$
        \mathrm{L}_Y\eta=0
        \quad ,\quad \mathrm{L}_Y\mathcal{H}=0 \ .
    $$
\end{itemize}
    
\end{dfn}
 Every (infinitesimal) contact symmetry preserves the Reeb vector field; that is, $\Phi^*\mathcal{R}=\mathcal{R}$ (or $[Y,\mathcal{R}]=0$). We have that an (infinitessimal) contact symmetry is an (infinitessimal) dynamical symmetry.


Associated with symmetries of contact Hamiltonian systems are
the concepts of {\sl dissipated} and {\sl conserved quantities}:

\begin{dfn}\label{dfn:diss-quan}A function $F\in\mathrm{C}^\infty(M)$ is:
\begin{itemize}

    \item A \textbf{conserved quantity} of a contact Hamiltonian system if
    $\mathrm{L}_{X_\mathcal{H}}F=0$ \ .
    \item A \textbf{dissipated quantity} of a contact Hamiltonian system if
    $
    \mathrm{L}_{X_\mathcal{H}}F=-(\mathrm{L}_\mathcal{R} \mathcal{H} )\,F \ .
    $
\end{itemize}
\end{dfn}

For contact Hamiltonian systems, symmetries are associated with dissipated quantities as follows:

\begin{thm} 
\label{th:dissipation}
{\rm (Dissipation theorem).} 
If $Y$ is an infinitesimal dynamical symmetry, 
then $F=-\mathrm{i}(Y)\eta$ is a dissipated quantity.
\end{thm}

In particular, the Hamiltonian vector field $X_\mathcal{H}$ is
trivially a dynamical symmetry and its dissipated quantity is the energy, 
$F=-\mathrm{i}(X_\mathcal{H})\eta=\mathcal{H}$;
that is: $\mathrm{L}_{X_\mathcal{H}}\mathcal{H}=-(\mathrm{L}_\mathcal{R} \mathcal{H} )\,\mathcal{H}$.

The Dissipation theorem is similar to the classical Noether's theorem. The converse of this result, that is, if every dissipated quantity is associated to an infinitesimal dynamical symmetry, is not true in general. Nevertheless, we can characterize them as follows: for any function $F$, we have an associated vector field: $F=-\mathrm{i}(Y_F)\eta$, namely $Y_F=-F\mathcal{R}$. Then, the results follows using a theorem in \cite{LeonLainz2}:
\begin{thm}
Let $X$ be a vector field on $M$. Then $\mathrm{i}(X)\eta$ is a dissipated quantity if, and only if, $\mathrm{i}([X,X_{\mathcal{H}}])\eta=0$.
\end{thm}

Every dissipated quantity changes with the same rate ($-\mathcal{R}(\mathcal{H})$), 
which suggests that the quotient of two dissipated quantities should be a conserved quantity. 
Indeed:
\begin{prop}
\label{prop:disscon} Given two functions $F_1,\,F_2\in\mathrm{C}^\infty(M)$:
\begin{itemize} 
    \item If $F_1$ and $F_2$ are dissipated quantities and $F_2\neq0$, then $F_1/F_2$ is a conserved quantity.
    \item If $F_1$ is a dissipated quantity and $F_2$ is a conserved quantity, then $F_1G_2$ is a dissipated quantity.
\end{itemize}
\end{prop}

If $H\neq0$, it is possible to assign a conserved quantity to an infinitesimal dynamical symmetry~$Y$. 
Indeed, from Theorem \ref{th:dissipation} and Proposition \ref{prop:disscon}, 
the function $-i(Y)\eta/H$ is a conserved quantity. 

Finally, contact symmetries can be used to generate new dissipated quantities 
from a given dissipated quantity.
In fact, as a corollary of definitions 
\ref{dfn:diss-quan} and \ref{dfn:sym} we obtain:

\begin{prop}
    If $\Phi\colon M\rightarrow M$ is a contact symmetry and 
    $F\colon M\rightarrow \mathbb{R}$ is a dissipated quantity, then so is $\Phi^*F$.
\end{prop}

\section{Example: Motion in a  gravitational field with friction}

Consider the motion of a particle in a vertical plane under the action of constant gravity, with friction proportional to velocity. The Hamiltonian system is $(M,\eta,\mathcal{H})$, where $M=\mathrm{T}^*\mathbb{R}^2\times\mathbb{R}$. Considering coordinates $(x,y,p_x,p_y,s)$ we have that:
$$
\eta=\mathrm{d} s - p_x\mathrm{d} x - p_y\mathrm{d} y\,,\quad \mathcal{H}=\frac{1}{2}\frac{p_x^2+p_y^2}{m}+mgy+\gamma s\,.
$$

The contact Hamiltonian vector field is
\begin{equation*}
X_\mathcal{H} = \Big(\frac{1}{2}\frac{p_x^2+p_y^2}{m}-mgy-\gamma s\Big)\frac{\partial}{\partial s} + \frac{p_x}{m}\frac{\partial}{\partial x} + \frac{p_y}{m}\frac{\partial}{\partial y} -
\gamma p_x\frac{\partial}{\partial p_x}-(mg+\gamma p_y)\frac{\partial}{\partial p_y}\ .
\end{equation*}
Which leads to the following equations for curves:
\begin{equation*}
    \ddot{x} + \gamma \dot{x} = 0  \quad,\quad
    \ddot{y} + \gamma \dot{y} + g = 0  \quad,\quad
    \dot{s} = \frac{1}{2}\frac{p_x^2+p_y^2}{m}-mgy-\gamma s \ .
\end{equation*}

One can check that the Energy is dissipated: \quad
$\mathrm{L}_{X_\mathcal{H}}\mathcal{H}=-\gamma\mathcal{H}$. Moreover, we have that $\frac{\partial \mathcal{H}}{\partial x}=0$, thus 
$\frac{\partial }{\partial x}$ is a contact symmetry. $p_x$ is the associated dissipated quantity and $\mathrm{L}_{X_\mathcal{H}}p_x=-\gamma p_x \ .$
Finally, we have the following conserved quantity:\
$\displaystyle\mathcal{H}/p_x=
(mp^2/2 +mgy + \gamma s)/p_x \, .$
\section*{Acknowledgements}
We acknowledge the financial support from the 
Spanish Ministerio de Ciencia, Innovaci\'on y Universidades project
PGC2018-098265-B-C33
and the Secretary of University and Research of the Ministry of Business and Knowledge of
the Catalan Government project
2017--SGR--932.

\end{document}